\documentstyle[12pt]{article}
\newcommand{\be}{\begin{equation}}
\newcommand{\nn}{\nonumber}
\newcommand{\bea}{\begin{eqnarray}}
\newcommand{\eea}{\end{eqnarray}}
\newcommand{\ba}{\begin{array}}
\newcommand{\ea}{\end{array}}
\newcommand{\ee}{\end{equation}}

\expandafter\ifx\csname mathbbm\endcsname\relax

\else

\fi \textheight 22cm \textwidth 15cm \topmargin 1mm \oddsidemargin
5mm \evensidemargin 5mm

\newcommand{\beas}{\begin{eqnarray*}}
\newcommand{\eeas}{\end{eqnarray*}}
\newcommand{\bes}{\begin{equation*}}
\newcommand{\ees}{\end{equation*}}
\newcommand{\dir}{\not\!\!{D}}

\newcommand{\lf}{\left}
\newcommand{\ri}{\right}
\newcommand{\f}{\frac}

\def\tr           {\mbox{\rm tr}\,}

\def\i2           {\mbox{$\frac{i}{2}$}}

\def\al           {\alpha}

\def\alb           {{\bar{\alpha}}}
\def\bet           {\beta}

\def\del           {\delta}

\def\ep           {\epsilon}

\def\ga           {\gamma}

\def\la           {\lambda}

\def\rh           {\rho}

\def\si           {\sigma}

\def\th{\theta}

\def\pl           {\partial}

\def\alb           {{\bar{\alpha}}}

\def\osi{{\bar {\si}}}

\def\thb        {{\bar {\theta}}}

\def\ola        {{\overline \la}}
\def\ow         {{\overline W}}

\begin{document}

\begin{titlepage}
\hfill \vbox{
    \halign{#\hfil         \cr
           hep-th/0308171 \cr
           IPM/P-2003/051 \cr
           } 
      }  
\vspace*{20mm}
\begin{center}
{\LARGE {  On Instantons and Zero Modes of ${\cal N}=1/2$ SYM Theory }\\
}

\vspace*{15mm} \vspace*{1mm} {Ali Imaanpur 
\footnote{Email: aimaanpu@theory.ipm.ac.ir}}

\vspace*{1cm}

{\it  Department of Physics, School of Sciences \\
Tarbiat Modares University, P.O. Box 14155-4838, Tehran, Iran\\
\vspace*{1mm}
Institute for Studies in Theoretical Physics and Mathematics (IPM)\\
P.O. Box 19395-5531, Tehran, Iran}\\
\vspace*{1cm}

\end{center}


\begin{abstract}
We study zero modes of ${\cal N}=1/2$ supersymmetric Yang-Mills action 
in the background of instantons. 
In this background, because of a quartic antichiral fermionic term in the 
action, the fermionic  solutions of the equations of motion are not in 
general zero modes of the action. Hence, when there are fermionic solutions, 
the action is no longer minimized by instantons. By deforming the instanton 
equation in the presence of fermions, we write down the zero modes equations. 
The solutions satisfy the equations of motion, and saturate the BPS bound. 
The deformed instanton equations imply that the finite action solutions have 
$U(1)$ connections which are not flat anymore.  

\end{abstract}

\end{titlepage}

\section{Introduction}
The duality correspondence between  supersymmetric gauge 
theories and matrix models \cite{VAFA, MAT} has had many interesting 
implications. Having originated from string theory, the duality with 
matrix models can be extended even beyond the known supersymmetric 
gauge theories. One such generalization comes by studying the D-branes in 
the background of 
graviphoton field \cite{GRASSI, CVAFA, KLEMM}. In this background, the spinor 
coordinates  
$\th^\al$ on the brane turn out to be nonanticommuting variables.  
And the effective theory on the brane comes out as an exotic supersymmetric 
gauge theory with nonanticommuting fermionic fields. In another approach, 
assuming that superspace coordinates do not anticommute, 
Seiberg \cite{SEI} has shown that one can still construct a super 
Yang-Mills Lagrangian which preserves only half of the supersymmetries. 
The renormalizability of this model has been studied in 
\cite{REY1, REY3, GRIS, BO, ROM, REY4, REY2}. While the generalization to 
${\cal N}=2$, along with other interesting aspects of noncommutative superspace
have further been explored in \cite{SEI2, ARAKI, FER, IVAN, TERA, REZA, CHA}.

The analysis of (anti)instantons in ${\cal N}=1/2$ SYM model parallels the one in 
ordinary  ${\cal N}=1$ SYM theory, though, in the former case the equation of motion  
for chiral fermions $\la$ becomes more involved when $F^+=0$. In this background, 
there are antichiral $\ola$ zero modes, which deform the equation for chiral 
fermions so that $\la$ cannot remain zero. Further, because of a quartic antichiral 
term in the action, the fermionic solutions to the eqautions of motion are not 
in general the zero modes of the action.  
This will have a further consequence that in the presence of 
fermions, instantons are not solutions to the equations of motion anymore. 
This can also be seen directly by noticing that in the presence of fermionic 
solutions (not the zero modes)  
the action has a value greater than the instanton charge, the difference being 
proportional to the deformation parameter $C^{\al\bet}$.      
In this note we would like to comment that in ${\cal N}=1/2$ $U(2)$ SYM model 
the instanton equation should be deformed as follows,  
\bea
&& F_{\mu\nu}^+ +\f{i}{2}C_{\mu\nu}\ola\ola =0 \label{ins1} \\
&& \dir \ola =0 \label{ins2} \, , 
\eea
where $F^+_{\mu\nu}$ is the self-dual part of the field strength, and $C_{\mu\nu}$ 
being the deformation parameter. 
The above equations, as we will see shortly, are also the zero mode equation 
for $\ola$. 
The solutions of the above equations, like instantons and their corresponding 
zero modes say in ${\cal N}=1$ theory, saturate a BPS bound in each topological 
sector specified by the instanton number $k$. However, unlike instantons in the 
presence of fermionic solutions in ${\cal N}=1/2$ theory, they do satisfy the 
equations of motion. This happens partly because in this configuration 
$\la =0$ is still a solution.
   
In what follows, we will see that for these states only the $U(1)$ part of 
the instanton equation is deformed and finite action solutions will have 
nonflat $U(1)$ connection. The $SU(2)$ instanton equation and the 
corresponding Dirac equation for the adjoint fermions, on the other hand, 
remain undeformed. This should be contrasted with the case of supersymmetric 
solutions (instantons) with fermions set to zero, where the finiteness of 
action requires one to consider only the flat $U(1)$ connections.     
In the case of 't Hooft one-instanton solution and the corresponding fermionic 
zero modes, we obtain the solutions of the $U(1)$ gauge fields. These, however, 
will have a zero $U(1)$ instanton number and thus belong to the $k=-1$ topological 
sector. This is in agreement with the fact that there are no $U(1)$ instantons 
except the flat ones, and that Eqs. (\ref{ins1}) and 
(\ref{ins2}) are the zero modes equations.

\section{Instantons and zero modes}

Let us begin with assuming that the superspace coordinates $\th^\al$ 
are not anticommuting, and instead they satisfy the following anticommutation 
relation
\be
\{\th^\al , \th^\bet \} = C^{\al\bet}\, , \label{DEF}
\ee
where $C^{\al\bet}$ is a constant and symmetric $2\times 2$ matrix. 
This deformation of the superspace has been studied earlier \cite{SCH, FERR, KLEMM}.  
The anticommutation relation (\ref{DEF}) will deform the supersymmetry algebra with ${\overline Q}^2$ 
proportional to the deformation parameter $C^{\al\bet}$.
Seiberg \cite{SEI} has considered the above deformation in  
${\cal N} =1$ supersymmetric model and has shown that half of the 
supersymmetries can be preserved. 
Indeed if $W^\al = (A_\mu , \la)$ denotes the usual ${\cal N}=1$ gauge super 
multiplet, then the Lagrangian of this ${\cal N}=1/2$ model 
 reads
\bea 
{\cal L} &=& i\tau \int d^2\th \tr W^\al W_\al -i{\bar \tau} \int d^2\thb 
\tr \ow^\alb\ow_\alb \\ \nn
&& + (i\tau -i{\bar \tau})\lf(-iC^{\mu\nu}\tr F_{\mu\nu}\ola\ola +\f{|C|^2}{4} 
\tr (\ola\ola)^2\ri)\, ,
\eea 
where
\[
C^{\mu\nu} \equiv C^{\al\bet}\ep_{\bet\ga}\si^{\mu\nu\ \ga}_\al \, 
\]
is a constant and antisymmetric self-dual matrix, and $|C|^2=C_{\mu\nu}C^{\mu\nu}$. 

The above Lagrangian is invariant under the following $Q$ deformed supersymmetry 
transformations, 
\bea
&& \del \la = i\ep D +\si^{\mu\nu}\ep \lf(F_{\mu\nu}+\f{i}{2}C_{\mu\nu}\ola\ola\ri)\\ \nn 
&& \del A_\mu =-i\ola\osi_\mu \ep \\ \nn  
&& \del D= -\ep\si^\mu D_\mu\ola \\ \nn
&& \del\ola =0  \nn \, ,
\eea
whereas ${\overline Q}$ is broken. A supersymmetric state is invariant under the 
above supersymmetric transformations. Setting $\la$ and $\ola$ to zero, with 
\be
F_{\mu\nu}^+ =0\, ,\label{INSTANTON}
\ee
gives such a state preserving the whole unbroken supersymmetry. This is of 
course also a solution to the equations of motion. Moreover, in ${\cal N}=1$ 
SYM theory, since in the 
background of instantons ($F^+=0$) there are antichiral zero modes with no chiral 
zero modes, instantons remain solutions 
to the equations of motion in the presence of fermionic solutions. 
But in ${\cal N}=1/2$ SYM theory this is not the story.

As mentioned in Introduction, in the background of instantons, and because the 
action of ${\cal N}=1/2$ SYM theory has a quartic antichiral fermionic term, 
the zero modes, in general, 
do not satisfy the equations of motion. Therefore, instantons will not remain 
solutions when there are fermions. To remedy this, we deform the instanton 
equation and show that they satisfy the equations of motion and like instantons 
have a finite action. The equations are
\bea
&& F_{\mu\nu}^+ +\f{i}{2}C_{\mu\nu}\ola\ola =0 \nn \\
&& \dir \ola =0 \nn \\
&& \la =0 \, . \label{BPS}
\eea
It is easy to see that a solution to (\ref{BPS}) is also a solution to the 
equations of motion. In fact, if we set $\la =0$, the equation of motion 
for the gauge fields is satisfied:
\bea
D_\mu \lf ( F^{\mu\nu} + iC^{\mu\nu}\ola\ola \ri) &=&
 D_\mu\lf( F^{\mu\nu -}+ F^{\mu\nu +} + \f{i}{2}C^{\mu\nu}\ola\ola   +\f{i}{2}C^{\mu\nu}\ola\ola\ri) \nn \\
&=& D_\mu\lf( F^{\mu\nu -} + \f{i}{2}C^{\mu\nu}\ola\ola \ri) \nn \\
&=& D_\mu\lf( F^{\mu\nu +} + \f{i}{2}C^{\mu\nu}\ola\ola \ri) =0 \, ,
\eea
where in the last equality we used the Bianchi identity
\be
D_\mu ( F^{\mu\nu +} - F^{\mu\nu -})=0 \, ,
\ee
with
\be
F_{\mu\nu}^{\pm} = \f{1}{2}(F_{\mu\nu}\pm \f{1}{2}\ep_{\mu\nu\rh\si}F^{\rh\si})\, .
\ee
The equations of motion for $\ola$ and $\la$ read
\bea
&& \overline {\dir} \la =-C^{\mu\nu} F_{\mu\nu}^+\ola- i\f{|C|^2}{2}(\ola\ola)\ola \nn \\
&& \dir \ola =0 \, .
\eea
which are also satisfied by solutions of (\ref{BPS}). 

Equivalently, one can check that the solutions to (\ref{BPS}) are absolute minima
of the action 
in each topological sector. To see this, let us again set $\la =0$, and add to the  
action a topological term proportional to the instanton number
\bea
S + k &=& 
\int d^4x\, \tr\!\!\lf( \f{1}{2}F^{\mu\nu}F_{\mu\nu} 
+iC^{\mu\nu} F_{\mu\nu}\ola\ola -\f{|C|^2}{4} (\ola\ola)^2  +\f{1}{4}\ep_{\mu\nu\rh\si}F^{\mu\nu}F^{\rh\si}\ri) \nn \\
&=& \int d^4x\, \tr\!\!\lf(  F^{\mu\nu +}F_{\mu\nu}^+  
+iC^{\mu\nu} F_{\mu\nu}\ola\ola -\f{|C|^2}{4} (\ola\ola)^2\ri) \nn \\
&=&  \int d^4x\, \tr\!\!\lf(F_{\mu\nu}^+ +\f{i}{2}C_{\mu\nu}\ola\ola \ri)^2 \geq 0 \, ,
\eea
henec for configurations satisfying (\ref{BPS}), the action is minimized and 
equal to $-k$. This also signals that fermions satifying (\ref{BPS}) 
are the true zero modes of the action.

For $U(1)$ gauge group, since $\ola$ is in the adjoint representation, the second 
equation of (\ref{BPS}) reduces to 
\be
\not\!{\pl}\ola =0
\ee
which has no normalizable solution on ${\bf R}^4$. Therefore in the $U(1)$ case, 
Eqs. (\ref {BPS}) reduce to the abelian instanton equations which are known  
to have no nontrivial solutions on ${\bf R}^4$ except the flat ones. 
However, for gauge groups of 
higher rank, there might be nontrivial solutions different than instantons. 
Let us hence consider the $U(2)$ gauge group. 

\section{Analysis of zero mode equations}

In this section we would like to analyse possible solutions to Eqs. (\ref{BPS}). 
We will see that in the background of fermionic solutions of Dirac equation, 
the $U(1)$ part of the connection can no longer be flat. Specially, we find 
solutions in the presence of adjoint fermions in the background 
of 't Hooft instantons. The $U(1)$ connections we find have a zero instanton 
number which is consistent with the fact that the deformed instanton equations 
are also the equations for zero modes.  

To begin with, let $T^a = (T^i =\f{\si^i}{2} , T^4=\f{1}{2})$, $i=1,2,3$ 
denote the generators of $SU(2)$ and $U(1)$ groups respectively. 
Now, to isolate the $SU(2)$ and $U(1)$ parts of Eqs. (\ref{BPS}), we expand the 
quadratic term $\ola\ola$ in that equation  
\bea
\ola\ola &=& \ola^a\ola^b T^a T^b = \f{1}{2}\ola^a\ola^b\lf( [T^a ,T^b] +
\{T^a ,T^b \}\ri) \nn\\ 
&=& \f{1}{2}\ola^i\ola^j( i\ep_{ijk}\, T^k + \f{\del_{ij}}{2}) 
+ \ola^i\ola^4\,  T^i + \f{1}{4}\ola^4\ola^4 \nn \\
&=& \f{1}{4}\ola^i\ola^i +\f{1}{4}\ola^4\ola^4+ \ola^i\ola^4 \, T^i   \, .
\eea 
So, for the $SU(2)$ part we have
\bea
&& F_{\mu\nu}^{+i} + \f{i}{2}C_{\mu\nu}\ola^i\ola^4  =0 \label{INST}\\
&& \dir \ola^i =0  \label{DIRAC}\, ,
\eea
while the $U(1)$ part reads
\bea
&& F_{\mu\nu}^{+4} +\f{i}{8}C_{\mu\nu}\ola^i\ola^i + \f{i}{8}C_{\mu\nu}\ola^4\ola^4  =0 
\label{MON} \\
&& \not\!{\pl}\ola^4 =0 \, .  
\eea
As said before, the last equation has no normalizable solution and thus we set 
$\ola^4 =0$. This will reduce Eqs. (\ref{INST}) and (\ref{DIRAC}) to the 
ordinary $SU(2)$ instanton 
equation and the corresponding Dirac equation for the adjoint antichiral fermions. 
Equation (\ref{MON}), however, reads
\be
F_{\mu\nu}^{+4} +\f{i}{8}C_{\mu\nu}\ola^i\ola^i =0 \, ,\label{MONOPOLE}
\ee
where $\ola^i$ are solutions to (\ref{DIRAC}). 
This is the equation we would like to further study. For this 
we consider the 't Hooft one instanton solution and the fermionic 
modes in the adjoint representation satisfying (\ref{DIRAC}). 

A solution to Dirac equation (\ref{DIRAC}) can in general be written in terms 
of ADHM data \cite{COR, WEI, DOREY}. In particular, in the background 
of one instantons one can write the explicit solutions \cite{COR, AMATI}. For the 
$SU(2)$ case, we quote the result for three different normalized 
solutions
\bea
\tr (\ola\ola) &=& \f{2}{\pi^2}\, \rh^2f(x)^3 \\
\tr (\ola\ola) &=& \f{6}{\pi^2} \, \rh^4 f(x)^4 \\
\tr (\ola\ola)  &=& \f{3}{\pi^2}\, \rh^2 r^2 f(x)^4 \, ,
\eea
where
\be
f(x)=\f{1}{r^2 +\rh^2}\, ,
\ee
and $r=\sqrt {\sum_{i=1}^4 x_i^2}$. Here $\rh$ is the instanton size, and 
we have set to zero the instanton position for simplicity.
The analysis of Eq. (\ref{MONOPOLE}) becomes easier if we make use of 
the Corrigan's identities
\bea
&& \rh^2 f(x)^3 = -\f{1}{8}\pl^\al\pl_\al f(x) \label{COR1} \\
&&  \rh^4 f(x)^4 = \f{1}{24} \pl^\al\pl_\al \lf( r^2f(x)^2 -2f(x) \ri) \label{COR2}\\
&&  r^2 \rh^2 f(x)^4 = -\f{1}{24} \pl^\al\pl_\al \lf( r^2f(x)^2 + f(x) \ri)\, .
\label{COR3} 
\eea 
Eq. (\ref{MONOPOLE}) now reads
\be
F_{\mu\nu}^{+4} =-\f{i}{8}C_{\mu\nu}\pl^\al\pl_\al K(x)\, ,
\ee
for $K(x)$ being any of the functions $\f{2}{\pi^2}f(x)\, , 
\f{6}{\pi^2}(r^2f(x)^2 -2f(x))$ or 
$\f{3}{\pi^2}(r^2f(x)^2 +f(x))$ appearing on the right hand sides of 
Eqs. (\ref{COR1}), (\ref{COR2}), and (\ref{COR3}), respectively.  
The solution to this equation 
-- up to a gauge transformation -- is then found to be
\be
A_\mu (x) = \f{i}{8}C_{\mu\nu}\pl_\nu K(x)\, ,\label{SOL}
\ee
which of course has a nonvanishing curvature. Therefore, what we have found is 
that in the presence of fermionic zero modes the $U(1)$ part of the connection 
cannot remain flat. This is in contrast with the supersymmetric solutions  
to the field equations (instantons) where one sets the fermions to zero, and for 
having a finite 
action solution one has to restrict to the flat part of the $U(1)$ connections. 

For instantons of higher topological charge, as mentioned earlier, one can 
write the adjoint zero mode solutions in terms of the ADHM data. Interestingly, 
using the Corrigan's identity, it can be seen that the expression for 
$\tr (\ola\ola)$ is actually a total derivative for all topological 
charges $k$,
\be
\tr (\ola\ola)_k = \pl^\al\pl_\al {\cal K}(x;k)\, ,
\ee
for some function ${\cal K}(x;k)$ which in turn has an explicit expression in 
terms of ADHM data \cite{COR, DOREY}.  
Therefore, we conclude that the solutions to (\ref{MONOPOLE}) for all values of 
$k$ will have the general form of the solution we found for $k=-1$ in (\ref{SOL}). 
Moreover, since ${\cal K}(x;k)$ has no singularity and at infinity goes like $r^{-2}$, 
the corresponding $U(1)$ gauge fields will all have zero instanton 
number.

\end{document}